\documentclass[%
 aip,
cp,  
 amsmath,amssymb,
 reprint,%
]{revtex4-2}

\usepackage{graphicx}
\usepackage{dcolumn}
\usepackage{bm}

\usepackage[utf8]{inputenc}
\usepackage[T1]{fontenc}
\usepackage{mathptmx}

\def\Eq#1{Eq.(\ref{#1})}

\def\Fig#1{Fig.~\ref{#1}}

\def\nd{\noindent}

\def\>{\rangle}
\def\<{\langle}

\def\ch{\text{ch}}

\def\adg{a^\dagger}
\def\Adg{A^\dagger}

\def\bdg{b^\dagger}

\def\dg{\dagger}

\def\w{\omega}
\def\del{\delta}
\def\gam{\gamma}

\def\Gam{\Gamma}
\def\al{\alpha}
\def\bt{\beta}
\def\sig{\sigma}

\def\eps{\epsilon}

\def\si{\text{s}}
\def\s{\text{s}}

\def\rad{\text{r}}
\def\st{\text{site}}

\def\th{\text{th}}
\def\cmi{\text{cm}^{-1}}
\def\psi{\text{ps}^{-1}}

\def\smn{\sideset{}{'}\sum_{\mu<\nu}}
\def\siL{\sum_{x=1}^\ell}
\def\siLm{\sum_{x=1}^{\ell-1}}
\def\smL{\sum_{\mu=1}^\ell}
\def\ssL{\sum_{\sig=1}^\ell}
\def\mn{{\mu\nu}}
\def\ms{{\mu\sig}}
\def\sn{{\sig\nu}}

\def\nb{\bar{n}}

\begin{document}

\title{Energy transfer in quantum molecular chain -- two models of inhomogeneity}

\author{B. A. Tay}
\email[Corresponding author: ]{BuangAnn.Tay@nottingham.edu.my}
\affiliation{Department of Foundation in Engineering, Faculty of Science and Engineering, University of Nottingham Malaysia, Jalan Broga, 43500 Semenyih, Selangor, Malaysia}

\date{\today}

\begin{abstract}
We study a linear chain of oscillators with inhomogeneity in their interactions with phonon bath. In a previous work on the Markovian master equation of the system, we investigated a model in which the difference in the site-phonon coupling between adjacent oscillators is the same throughout the chain. Here we look into another model in which the oscillators are coupled to the phonon bath with alternating strength at successive sites. Whereas in the first model all exciton modes are connected, in the second model they are coupled in pairs that are not connected to each other. Owing to this special structure in the coupling, the excitation numbers of different modes can be solved exactly in the steady state. In the first model, the minima of the excitation profile in the site basis occur at the edges of the chain, whereas in the second model the maxima occur at the edges. The energy transfer efficiency in the first model is affected by the source power whereas in the second model the efficiency is independent of it. A distinct feature in the second model is that a sink placed at the middle of the chain is able to distinguish between chains with even and odd number of sites. The energy transfer efficiency in a chain with even number of sites is higher than a chain with odd number of sites. Therefore, it reveals the discrete nature of the chain. In the limit of very long chain when the discreteness of the chain is less evident, the efficiencies approach each other.

\end{abstract}

\maketitle

\section{Motivations}
\label{SecIntro}

In this paper, we summarise our work on the energy transfer in quantum molecular chain \cite{Tay21} under the framework of open quantum systems \cite{Breuer,May11}, and introduce a new model of site-phonon coupling as a further application to the results. The Hamiltonian we study describes the transfer of energy in photosynthetic systems \cite{May11,Jang18}. It also contained solitonic solutions when applied to the energy transfer in $\alpha$-helix protein chain \cite{Davydov90}.
Owing to the complexity in real systems, studies are mostly carried out numerically with the approximation that each site contains a single excited level \cite{Mohseni08,Olaya08,Plenio08,IshizakiPNAS09,IshizakiJCP09a,Chin10,Plenio12,Jang18}. The time evolution of the excitation is followed and the energy yield is obtained.

To carry out an analytic study on the system, we study a simplified version of a linear chain of oscillators interacting with a phonon bath.
In our approach, we first simplify the Hamiltonian through a unitary transformation into a dressed basis.
A further transformation into an exciton basis then diagonalises the Hamiltonian of the chain.
Supplemented by a few assumptions on the system, we obtain a reduced dynamics of the chain in a phonon bath under thermal equilibrium.

In our previous studies \cite{Tay21}, we assumed a model in which a difference in the site-phonon coupling between nearest neighbours is the same. The relaxation rate could be calculated analytically and all the exciton modes are coupled.
We also studied the excitation profile of the chain and the energy transfer efficiency in the steady state in detail. Here, we explore a different model in which the oscillators are coupled to the phonon bath at alternating strength in the subsequent sites. This implies that the difference in the site-phonon coupling changes sign between pairs of subsequent nearest neighbours.
In this model, pairs of exciton modes are coupled separately. Each pair does not interact with the rest. The excitonic structure is very similar to a system that consists of many independently coupled two-level systems with different excitation energy.
This model can be solved analytically in the steady state. It leads to very different behaviours compared to the previous model.

In the following, we will first summarise the main steps taken and results obtained in our previous work.
We then calculate the relevant quantities of the second model and compare the results with the first model.
We hope that our analytic work on such simplified systems can provide insights into the energy transfer mechanism in quantum molecular chains.

\section{Equations of motion in excitation number}
\label{SecEqMotion}

In photosynthetic systems, light is absorbed by protein pigments held in a protein scaffold \cite{Adolphs06}. Light creates excitations that are transferred through the chlorophyll molecules in the protein pigments to reaction centers, where energy is trapped in a useful form of chemical compound. During the process, the excitations are influenced by the vibrations of the molecules in the protein scaffold, which is described by phonon fields. In the well-studied Fenna-Matthews-Olson (FMO) pigment protein complex in green sulphur bacteria, each of the complex consists of three monomers. Each monomer contains seven bacteriochlorophyll-a (BChla) molecules that are connected to each other in a complicated way.

To facilitate our analytical works, we consider a simplified version of a linear chain of coupled oscillators that interact with their nearest neighbours
\begin{align}   \label{Hch}
    H_0&= \siL \w_0 \adg_x a_x+J\siLm \big(\adg_x a_{x+1}+a_x\adg_{x+1}\big)\,,
\end{align}
where we use the units $\hbar=c=1$. $\adg_x, a_x$ are the creation and annihilation operators at site-$x$. The oscillators have a same natural frequency $\w_0$ and an intersite coupling $J$.
The chain lies in a protein scaffold whose vibrations are effectively described by a field of phonons $H_q=\sum_q \w_q \bdg_q b_q$. The phonons interact with the oscillators through the potential
\begin{align}   \label{V}
    V&= \siL \sum_q \w_q \chi^{(x)}_q \adg_x a_x \big( b_{-q}+\bdg_q\big)\,,
\end{align}
where $\chi^{(x)}_q$ is a real dimensionless coupling constant.
In most works exciton basis is introduced at this stage. However, in our approach we first introduce a new basis through the unitary transformation \cite{May11}
\begin{align}   \label{U1}
    U=\exp\bigg(-\sum_q \sum^\ell_{x=1}   \chi_q^{(x)} \adg_x a_x (b_{-q}-\bdg_q)\bigg) \,.
\end{align}
In the new basis, the number operator of the dressed oscillator remains the same as the number operator $\adg_x a_x$ of the bare oscillator \cite{Tay21}. This simplifies our discussion when we consider the excitation numbers of the oscillator since they are equivalent in both basis. The natural frequency is modified by the interaction between the oscillators and phonons. We assume that this modified term is negligible in order to diagonalize $H_0$.

\begin{figure}[t]
\centering
\includegraphics[width=5.in]{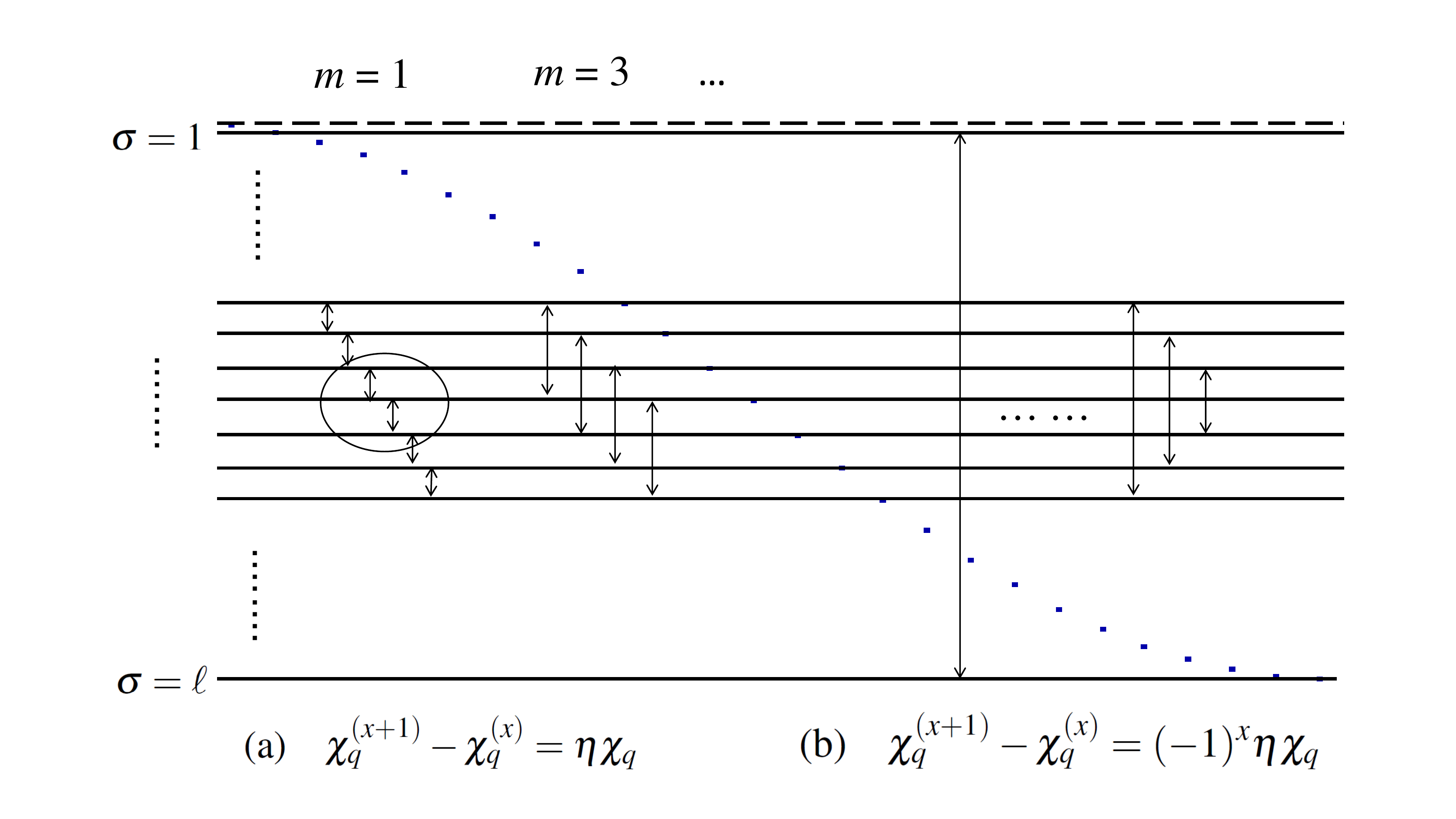}
\caption{Energy spectrum of excitons. Allowed transitions between excitons are shown for two models. (a) First model \eqref{chij+ve}, only transitions between exciton indices separated by $m=\nu-\mu=1, 3, \cdots$ are permitted. The fastest relaxation rate \eqref{gamc+} occurs to index lying in the middle of the set $m=1$ (circled in the figure). (b) Second model \eqref{chij-ve}, only transitions between two modes satisfying $\mu+\nu=\ell+1$ are allowed.  Relaxation rate is constant \eqref{gamc-}.
}
\label{fig1}
\end{figure}

The transformation introduces into the interaction a difference in the interaction $\chi^{(x)}_q$ between neighbouring sites with phonons
\begin{align}   \label{V'}
   V'= J \sum_q \siLm ( \chi_q^{(x+1)}-\chi_q^{(x)}) &(\adg_x a_{x+1}-\adg_{x+1} a_x)(b_{-q}-\bdg_q)\,.
\end{align}
This fact enables us to consider different models of inhomogeneity in $\chi_q^{(x)}$.
After this transformation, we introduce the exciton basis through a discrete sine transform \cite{May11}
\begin{align}   \label{A}
    A_\mu&\equiv\sqrt{\frac{2}{\ell+1}}\siL \sin\left(k_\mu x\right) \, a_x\,,\\
    k_\mu&\equiv \frac{\pi\mu}{\ell+1}\,,\label{k}
\end{align}
where $\mu=1,2,\cdots,\ell$.
The exciton basis diagonalizes the Hamiltonian of the chain into
\begin{align}   \label{H0A}
    H'_0&=\smL \w_\mu \Adg_\mu A_\mu\,,
\end{align}
with the exciton energy
\begin{align}   \label{wb}
    \w_\mu&\equiv \w_0+2J\cos k_\mu \,,
\end{align}
assuming that the correction to $\w_0$ is negligible.
The spectrum is modulated by the cosine function as shown in \Fig{fig1}.
The advantage of the basis \eqref{A} is that the exciton energy is a monotonically decreasing function of $\mu$.
Owing to this fact, we can interpret the relaxation process among the excitons in a straight-forward way in the next section.

As already explained in Ref.~\cite{May11}, there are two modes of energy transfer in quantum molecular chains. The first mode occurs through excitation hopping from one site to the next. The second mode involves transfer of energy between delocalised states that extend over several sites in the form of wave packets. The second mode is the dominant mode of transfer when intramolecular relaxations occur at a slower rate than intermolecular transitions. It is the second mode of transfer that we try to describe through the exciton basis.

In the exciton basis, the interaction Hamiltonian takes the form
\begin{align}   \label{V'exc}
    V'&=J \sum_q \smn \sum_{x=1}^{\ell-1} \big(\chi_q^{(x+1)}-\chi_q^{(x)}\big)   c_\mn(x) (L^\dg_\mn-L_\mn)( b_{-q} -\bdg_q) \,,
\end{align}
where $L_\mn ^\dg\equiv \Adg_\mu A_\nu$ is an exciton raising operator. $c_\mn(x)$ is a coefficient whose sum over $x$ can be calculated explicitly. We will give its expression in the next section.

We assume that the difference in the inhomogeneity is sufficiently small that it can be used as a perturbation parameter to obtain the Markovian master equation of the chain using standard method. The phonon field is assumed to be in thermal equilibrium. The average number of excitation in the $\sig$-mode is given by $n_\sig\equiv\<A^\dg_\sig A_\sig\>$, where $\<\cdot\>$ denotes a trace over the density matrix of the exciton. The equation of motion of $n_\sig$ can be worked out to be
\begin{align}   \label{ndt}
    \frac{dn_\sig}{dt}\bigg|_\ch&=
    \sum_{\mu=1}^{\sig-1}\Gam_\ms\big[n^\th_\ms(n_\mu-n_\sig)
            + n_\mu(1+n_\sig)\big]
    +\sum_{\nu=\sig+1}^\ell\Gam_\sn\big[n^\th_\sn(n_\nu-n_\sig)
    -n_\sig(1+ n_\nu)\big] \,.
\end{align}
We will present the expression of the relaxation rate of exciton $\Gam_\mn$ in the next section.
The phonon bath has the canonical distribution
\begin{align} \label{nbmn}
        n^\th_\mn&\equiv \frac{1}{e^{\bt \w_\mn}-1}\,,
\end{align}
with inverse temperature $\bt=1/(k_B T)$. The energy gap between two exciton modes is $\w_\mn\equiv\w_\mu-\w_\nu$. \Eq{ndt} is a set of coupled nonlinear equations in $n_\sig$.
We note that in the steady state, the off-diagonal components of the excitations vanish \cite{Tay21}.

Assuming that there is a constant source supplying a rate of energy $s$ at site-1, the $\sig$-mode exciton gains a portion of power
\begin{align}   \label{s1}
    s_\sig\equiv\al_\sig^{(1)}s
\end{align}
from the source, where (with $z=1$ in \Eq{s1})
\begin{align} \label{alz}
    \al_\sig^{(z)}=\frac{2}{\ell+1} \sin^2(k_\sig z)
\end{align}
is a weight factor $\sum_\sig \al_\sig^{(z)}=1$. Later on in our discussion, we denote the site where the sink is attached to by $z$.

We consider two types of loss mechanisms. One of them is the loss to radiation, where an exciton relaxes to the ground state of the system and radiates off a photon. We assume that the rate of radiation $\gam_\rad$ at each site is the same. The other loss mechanism is absorption of excitons by the sink. The energy absorbed turns into useful energy. If the sink is prepared at site-$z$, then each mode will experience a loss rate of $\al_\sig^{(z)}\gam_\si$ to the sink. Effects of the interactions between the chain and phonon bath are already included in \Eq{ndt}.

After including all the effects, we obtain the following set of rate equations for the number of excitations
\begin{align}   \label{dndttot}
        \frac{dn_\sig}{dt}
        &=s_\sig- \eps^{(z)}_\sig n_\sig
        +\frac{dn_\sig}{dt}\bigg|_\ch\,, \qquad  \sig=1,2,3,\cdots,\ell\,,\\
         \eps^{(z)}_\sig&\equiv \al^{(z)}_\sig \gam_\si+\gam_\rad\,, \label{eps}
\end{align}
where the sink is prepared at site-$z$.
We denote the excitation number in the steady state by $\nb_\sig$.
When we sum \Eq{dndttot} over all the modes in the steady state, we obtain a consistency equation
\begin{align}   \label{sumnt}
    s&=\ssL\eps^{(z)}_\sig \nb_\sig\,.
\end{align}

There are two clearly separated time scales in the reduced dynamics. During the shorter time scale of the order of $1/\Gam_\mn$, the excitations will achieve approximate equilibrium before the longer one of the order of $1/\eps^{(z)}_\sig$ sets in as the system approaches the steady state.

This set of equation is similar to those obtained in in biological systems \cite{Frohlich68,Frohlich68b} in an attempt to explain energy storage and energy transfer. It was found that when the source power exceeds certain limit, the excitations accumulate to the lowest exciton mode.
An example of this collective oscillation to the lowest mode was reported recently in protein macromolecules \cite{Nardecchia18}.

\section{Two models of inhomogeneity in site-phonon coupling}
\label{SecTwoModels}

\noindent \textit{(1) First model}

\hfill

In the first model of inhomogeneity in site-phonon coupling, we assume that the difference between consecutive sites is constant along the chain
\begin{align}   \label{chij+ve}
    \chi_q^{(x+1)}-\chi_q^{(x)}&= \eta \chi_q\,, \qquad x=1,2,\cdots,\ell-1\,,
\end{align}
where $\eta$ is a parameter that estimates the magnitude of the difference. We assume that it is a small parameter and use it as a perturbation parameter to obtain the Markovian master equation.
The sum of the coefficient $c_\mn(x)$ over $x$ in \Eq{V'} can then be calculated analytically to yield \cite{Tay21}
\begin{align}   \label{cmn+}
    \sum_{x=1}^{\ell-1}c_{\mu\nu}(x)&=\frac{4}{\ell+1} \frac{\sin k_\nu\sin k_\mu}{\cos k_\nu-\cos k_\mu}\,,  \qquad\nu-\mu=1,3,5,\cdots,
\end{align}
and 0 otherwise.
An interesting feature of this model is that only those modes satisfying the condition $\nu-\mu=$ odd are coupled, see \Fig{fig1}(a) for illustrations.
It follows that the relaxation rate of excitons between two modes is \cite{Tay21}
\begin{align}   \label{gamc+}
        \Gam_\mn&=\frac{4\eta^2 \gam_\text{d}}{(\ell+1)^2}\frac{\sin^2k_\mu \sin^2k_\nu}{(\cos k_\mu-\cos k_\nu)^4}\,,
        \qquad \nu-\mu=1,3,5,\cdots,
\end{align}
and 0 otherwise, where $\gam_\text{d}$ is the dephasing rate of a single site under the influence of a phonon bath.
Notice that the denominator is proportional to the fourth-order of the energy gap between two excitons.
Therefore, the larger the difference in the index $m=\nu-\mu$, the wider the energy gap (cf.~\Fig{fig1}) and the smaller the relaxation rate. We can then group the transitions into sets labelled by $m$, as illustrated in \Fig{fig1}(a).
The transitions between adjacent levels labeled by $m=1$ are fastest.
The square of sine functions in the numerator in \Eq{gamc+} indicates that the rate is fastest for excitons that lie at the center of the spectrum, highlighted by the circle in \Fig{fig1}(a).
Based on this information, we infer that the excitons relax by cascading down the energy spectrum. The lowest level of the excitons then acts as a metastable state before the excitation finally relaxes to the true ground state by losing energy to the sink or radiation.
An analysis shows that the relaxation rate increases as $\ell^2$ in the long chain limit \cite{Tay21}.

\hfill

\noindent \textit{(2) Second model}

\hfill

In the second model, we assume that the sites interact with the phonon bath in alternating strength, depending on whether they are odd or even sites in the chain. Consequently, we have the following difference in the site-phonon coupling
\begin{align}   \label{chij-ve}
    \chi_q^{(x+1)}-\chi_q^{(x)}&=(-1)^x \eta \chi_q\,, \qquad x=1,2,\cdots,\ell-1\,.
\end{align}
It leads to the coefficient
\begin{align}   \label{cmn-}
    \sum_{x=1}^{\ell-1}(-1)^x c_{\mu\nu}(x)&=\cos k_\nu-\cos k_\mu\,,  \qquad\mu+\nu=\ell+1 \text{ only},
\end{align}
and 0 otherwise. Notice that in this model only two modes satisfying the condition $\mu+\nu=\ell+1$ are coupled.
Examples of energy levels coupled in pairs are illustrated in \Fig{fig1}(b).
The relaxation rate turns out to be constant for all pairs,
\begin{align}   \label{gamc-}
        \Gam_2&\equiv\Gam_\mn=\frac{1}{4}\eta^2 \gam_\text{d}\,,
        \qquad \mu+\nu=\ell+1 \text{ only},
\end{align}
and 0 otherwise. It suggests that the system behaves like a set of independent two-level systems (with different energy gaps) separately connected to the source, phonon bath and sink. In this case, the set of equations \eqref{dndttot} together with \Eq{ndt} reduces to a set of equations coupled in pair,
\begin{align}   \label{dndttot2lev}
        \frac{dn_\mu}{dt}
        &=s_\mu- \eps^{(z)}_\mu n_\mu
        +\Gam_2 \left[n^\th_\mn(n_\nu-n_\mu)-n_\mu(1+n_\nu)\right]\,,\\
        \frac{dn_\nu}{dt}
        &=s_\nu- \eps^{(z)}_\nu n_\nu
        -\Gam_2 \left[n^\th_\mn(n_\nu-n_\mu)-n_\mu(1+n_\nu)\right]\,,\label{dndttot2levb}
\end{align}
where $\mu+\nu=\ell+1$ only.
For odd $\ell$, the unpaired mode $(\ell+1)/2$ evolves alone, i.e., it is not coupled to other modes,
\begin{align}
    \frac{dn_{(\ell+1)/2}}{dt}&=s_{(\ell+1)/2}- \eps^{(z)}_{(\ell+1)/2} n_{(\ell+1)/2}\,,\qquad \text{ for odd $\ell$ only.}
\end{align}
Denoting $\nb$ as the excitation number at the steady state, the total excitation number of two coupled modes is
\begin{align}   \label{N2mu}
        \nb_\mu+\nb_\nu&=2 \frac{s_\mu}{\eps^{(z)}_\mu}\,,\qquad \mu+\nu=\ell+1 \text{ only}\,,\\
        \nb_{(\ell+1)/2}&=\frac{s_{(\ell+1)/2}}{\eps^{(z)}_{(\ell+1)/2}}\,, \qquad \text{for odd $\ell$ only.}
\end{align}
To obtain \Eq{N2mu}, we use $s_\mu=s_\nu$ and $\eps^{(z)}_\mu=\eps^{(z)}_\nu$ since $\al^{(z)}_\mu=\al^{(z)}_\nu$ for $\mu+\nu=\ell+1$.
Owing to the simplicity in the coupling, the model has exact solution in the steady state. Using \Eq{N2mu}, we solve the simultaneous equations \eqref{dndttot2lev}-\eqref{dndttot2levb} at the steady state to get
\begin{align}   \label{n2mu}
        \nb_\mu&=\frac{1}{2}
        \left(\frac{\eps^{(z)}}{\Gam_2}+2\frac{s_\mu}{\eps_\mu}+2n^\th_\mn +1\right)- \sqrt{\frac{1}{4}\left(\frac{\eps^{(z)}}{\Gam_2}
        +2\frac{s_\mu}{\eps_\mu}+2n^\th_\mn +1\right)^2-\frac{s_\mu}{\Gam_2}-2n^\th_\mn \frac{s_\mu}{\eps_\mu}}\,,\qquad \mu+\nu=\ell+1 \text{ only}\,.
\end{align}
In \Eq{n2mu}, we have chosen the negative branch of the square root, otherwise $\nb_\nu$ can be negative, which is not permitted.

\section{Excitation profiles in the site basis}
\label{SecProfile}

\begin{figure}[t]
\centering
\includegraphics[width=6.in,trim={4cm 12cm 2cm 11cm}]{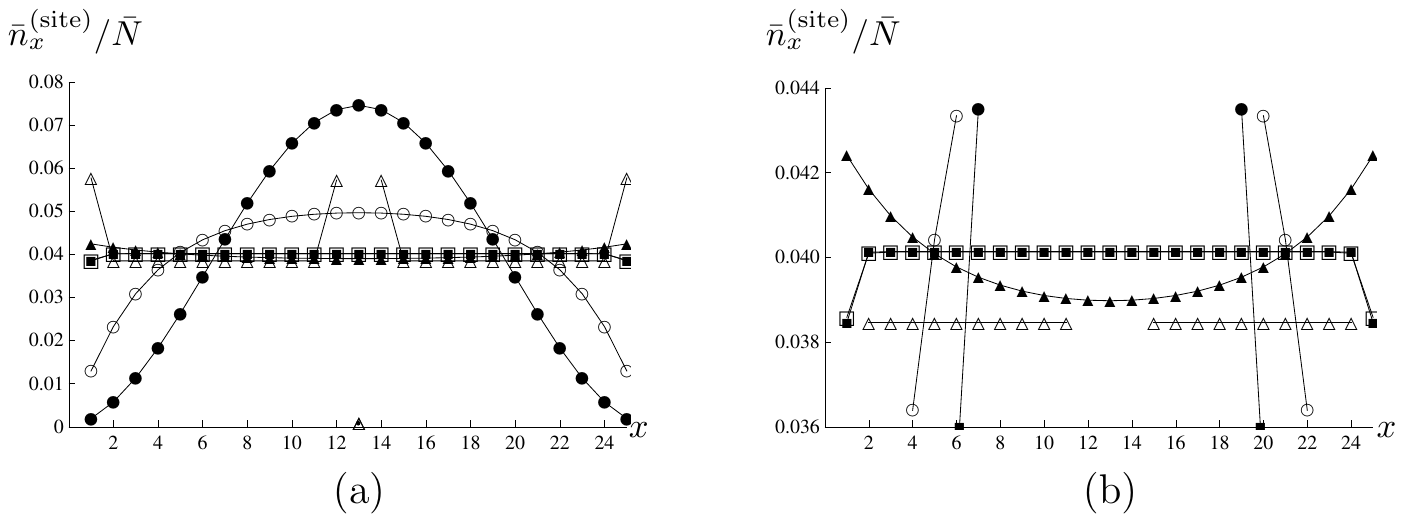}
\caption{Normalized excitation population in the site basis along a chain with 25 sites.
Filled (empty) shapes denote excitation profiles with a sink positioned at the end (middle) of the chain. Two values of source power are plotted. For low source power $s=0.001\,\psi$, $(\blacksquare,\square)$ label profiles in the first model (they almost overlap), whereas $(\blacktriangle,\vartriangle)$ label profiles in the second model. For high power source $s=10\, \psi$, only profiles in the first model $(\bullet,\circ)$ are shown. \Fig{fig2}(a) shows the whole range of excitation profile.
\Fig{fig2}(b) is a magnified view of \Fig{fig2}(a) in the range $0.036\leq \nb^{(\st)}_x/N\leq 0.044$.
}
\label{fig2}
\end{figure}

The excitation number at site-$x$, $\nb^{(\st)}_x\equiv\<\adg_x a_x\>$, can be obtained by summing over all the exciton modes weighted by $\al^{(x)}_\mu$ \cite{Tay21}
\begin{align}   \label{nsite}
    \nb^{(\st)}_x&=\smL
        \al^{(x)}_\mu \nb_\mu\,.
\end{align}
Let us compare the excitation profiles of the two models using typical parameters from the well-studied Fenna-Matthews-Olson (FMO) pigment protein complex in green sulphur bacteria \cite{Adolphs06,Mohseni08,Tay21}. We have approximately $J=100\,\cmi$ and $\w_0=12,500 \,\cmi$. For decay rate, we use the dephasing rate $\gam_\text{d}=20\,\psi$, radiation rate $\gam_\rad=0.001\text{ ps}^{-1}$, and a trapping power of the sink $\gam_\s=1\,\psi$. We choose a source power of one excitation per nanosecond $s=0.001\,\psi$ as a reference of low power source, and an inhomogeneity parameter of $\eta=0.1$. We start with a chain at the ground state as an initial condition.

\Fig{fig2} plots the excitation profiles in the site basis along a chain with 25 sites. Profiles of chain with a sink positioned at the end or middle of the chain are labelled by filled shapes $(\blacksquare, \bullet, \blacktriangle)$ or empty shapes $(\square, \circ, \vartriangle)$, respectively.

\hfill

\nd \textit{(1) First model}

\hfill

The excitation profile in the site basis \eqref{nsite} and the efficiency of the first model had been studied in detail in Ref.~\cite{Tay21}. In this model, the position of the sink does not affect the overall feature of the profile much, see \Fig{fig2}(a) for examples of low source power $s=0.001\,\psi$, where the excitation profiles are labelled by the $\blacksquare$- and $\square$-curves.
We notice that the excitations distribute almost uniformly across the chain and the two curves almost coincide. \Fig{fig2}(b) gives a magnified view of the profiles. The minima of the excitations occur at the edges of the chain.

When the power of the source exceeds certain limit, the excitations will ``condense'' to the lowest mode \cite{Frohlich68,Frohlich68b}. Now the excitation profile in the site basis has a prominent maximum at the middle of the chain, illustrated in \Fig{fig2}(a) by the $\bullet$- and $\circ$-curves for a source power of $s=10\,\psi$.
Intuitively, we expect that positioning the sink at the middle of the chain will lead to a higher energy transfer efficiency. This is confirmed by the results in the next section.

\hfill

\nd \textit{(2) Second model}

\hfill

In contrast to the first model, the excitation profiles in the second model are independent of the source power. With the help of \Eq{N2mu}, we can see this by casting \Eq{nsite} in a form proportional to the source power $s$,
\begin{align}   \label{nsite2}
    \nb^{(\st)}_x &=\sum_{\mu=1}^\text{mid}
        \al^{(x)}_\mu (\nb_\mu+\nb_\nu)
        = f(x,z)\cdot s\,,
\end{align}
where we use $\al^{(x)}_\nu=\al^{(x)}_\mu$ for $\mu+\nu=\ell+1$.
The symbol ``mid'' equals $\ell/2$ or $(\ell+1)/2$ depending on whether $\ell$ is even or odd, respectively. The profile function
\begin{align}   \label{f}
    f(x,z)&\equiv \sum_{\mu=1}^\text{mid} \al^{(x)}_\mu \frac{\al^{(1)}_\mu}{\eps^{(z)}_\mu}
    \big(2-\del_{\mu,(\ell+1)/2}\big)
\end{align}
is independent of $s$. It is determined by the position of the sink $z$. Since the total excitation number $\bar{N}$ is also proportional to $s$, the normalize excitation population $\nb^{(\st)}_x/\bar{N}$ of the second model in the site basis is therefore independent of $s$. This is a special feature of the second model where only exciton modes within each coupled pair are connected. In sharp contrast, in the first model all the exciton modes are coupled.

Unlike the first model, in the second model the position of the sink affects the profile of excitations greatly.
The detail of the excitation profile is determined mostly by the distribution of the sink power $\al^{(z)}_\sig\gam_\s$ across the exciton modes, cf.~\Eq{eps}. For a sink positioned at the end of the chain, excitons closer to the middle of the energy spectrum experience greater loss to the sink where $\al^{(\ell)}_\sigma$ is greatest. When the contribution of all the modes are summed up, the maxima of the excitation profile in the site basis occur at the edges of the chain, as illustrated by the $\blacktriangle$-curve in \Fig{fig2}(b). This is in sharp contrast to the first model where minima occur at the edges of the chain.

\begin{figure}[t]
\centering
\includegraphics[width=3.2in,trim={4cm 10cm 2cm 11cm}]{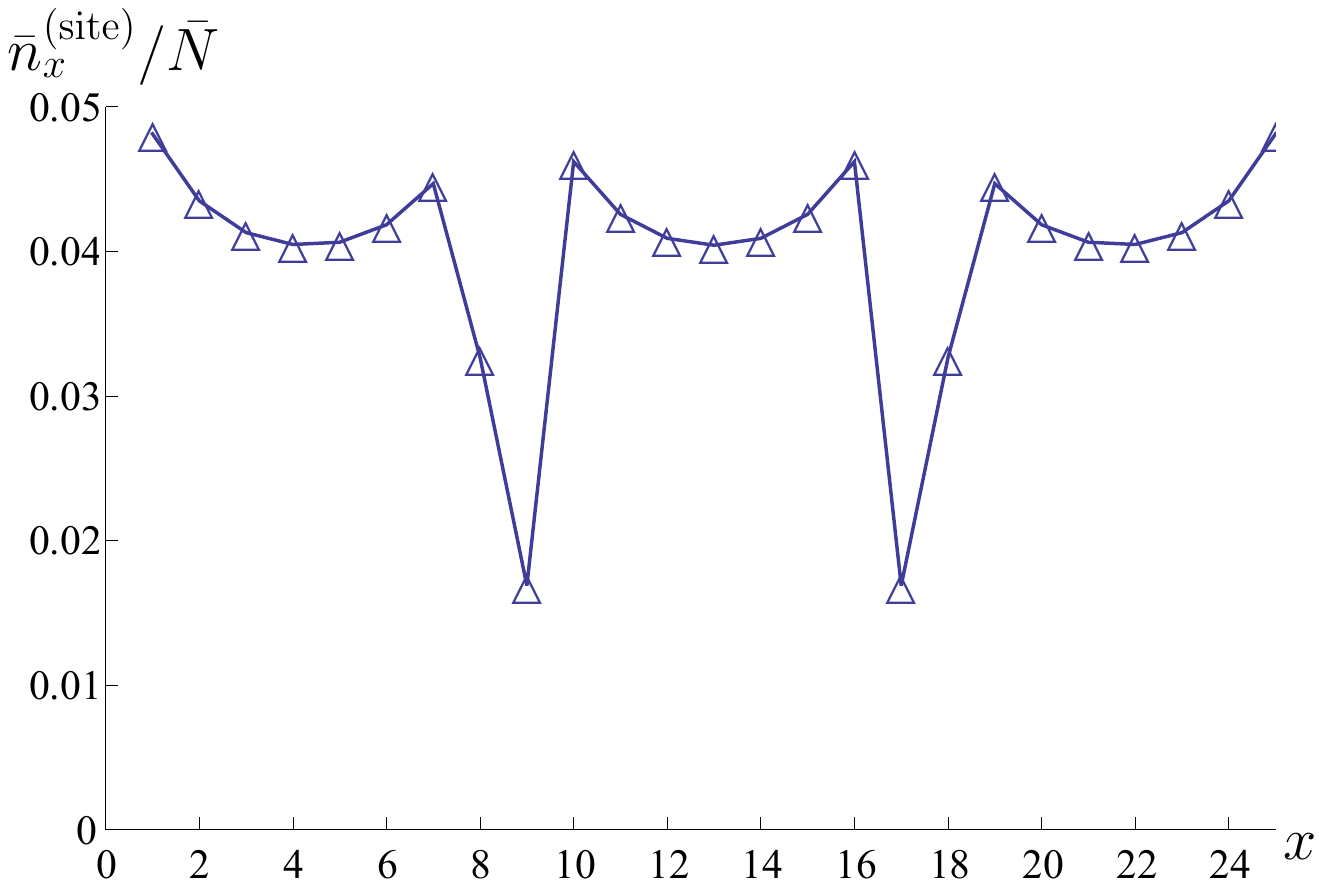}
\caption{Excitation profile in the site basis for a 25-site chain with a sink placed at a distance of three-quarter ($z=17$) from the origin. The same parameters with those of the $\vartriangle$-curve in \Fig{fig2} are used.
}
\label{fig3}
\end{figure}

When the sink is shifted to the middle of the chain, the sink power $\al^{(z)}_\sig\gam_\si$ vanishes at alternate exciton indices for a chain with odd number of sites. This means that half of the modes lose energy through radiation (and the phonon bath) only. As a result, the efficiency is expected to be lower, see the discussion in the next section. It turns out that the profile in the site basis exhibits maxima at the edges too, see the $\vartriangle$-curve in \Fig{fig2}(a).
Most interestingly, the middle site-13 has a sharp minimum with almost zero excitation.
On either side of the sharp minimum, we find two hollow excitation profiles.
In the case of even $\ell$, the profile is similar to odd $\ell$ except now there are two minima at the middle of the chain (not shown in the figure). By carefully arranging the sink at a distance of three quarter $z=17$ from the edge of the chain, it is possible to create an excitation profile with three hollow structures, as shown in \Fig{fig3}.

\section{Energy transfer efficiency}
\label{SecEff}

The efficiency of energy transfer along the chain is defined to be the ratio of the rate of useful energy absorbed by the sink to the rate of total energy supplied by the source
\begin{align}   \label{etaEsumsig}
       \eta_e&=\frac{1}{s} \ssL \w_\sig\gam_\s \al^{(z)}_\sig \nb_\sig\,.
\end{align}
We use the fact that a $\sig$-mode exciton loses $\w_\sig$ of energy to the sink at a rate of $\gam_\si\al^{(z)}_\sig\nb_\sig$ when it relaxes to the ground state.

We now summarise the results of the energy transfer efficiency of the first model already studied extensively in Ref.~\cite{Tay21}. In general, the efficiency reduces as the length of the chain increases.
In Ref.~\cite{Tay21}, it was shown that the external parameters, such as $\gam_\si$ and $\gam_\rad$, play a dominant role in determining the energy transfer efficiency. Internal parameters related to the chain, such as $J, \chi^{(x)}_q, \w_0$ and model related parameters such as $\eta$, have mild influence on the efficiency.

The analytical expressions of energy transfer efficiency for low power and high power sources were already obtained in Ref.~\cite{Tay21}. In general, the efficiency depends on the position of the sink. Below we study the efficiency for two positions of the sink, either it lies at the end or at the middle of the chain.

\begin{figure}[t]
\centering
\includegraphics[width=3.6in,trim={4cm 9cm 2cm 11cm}]{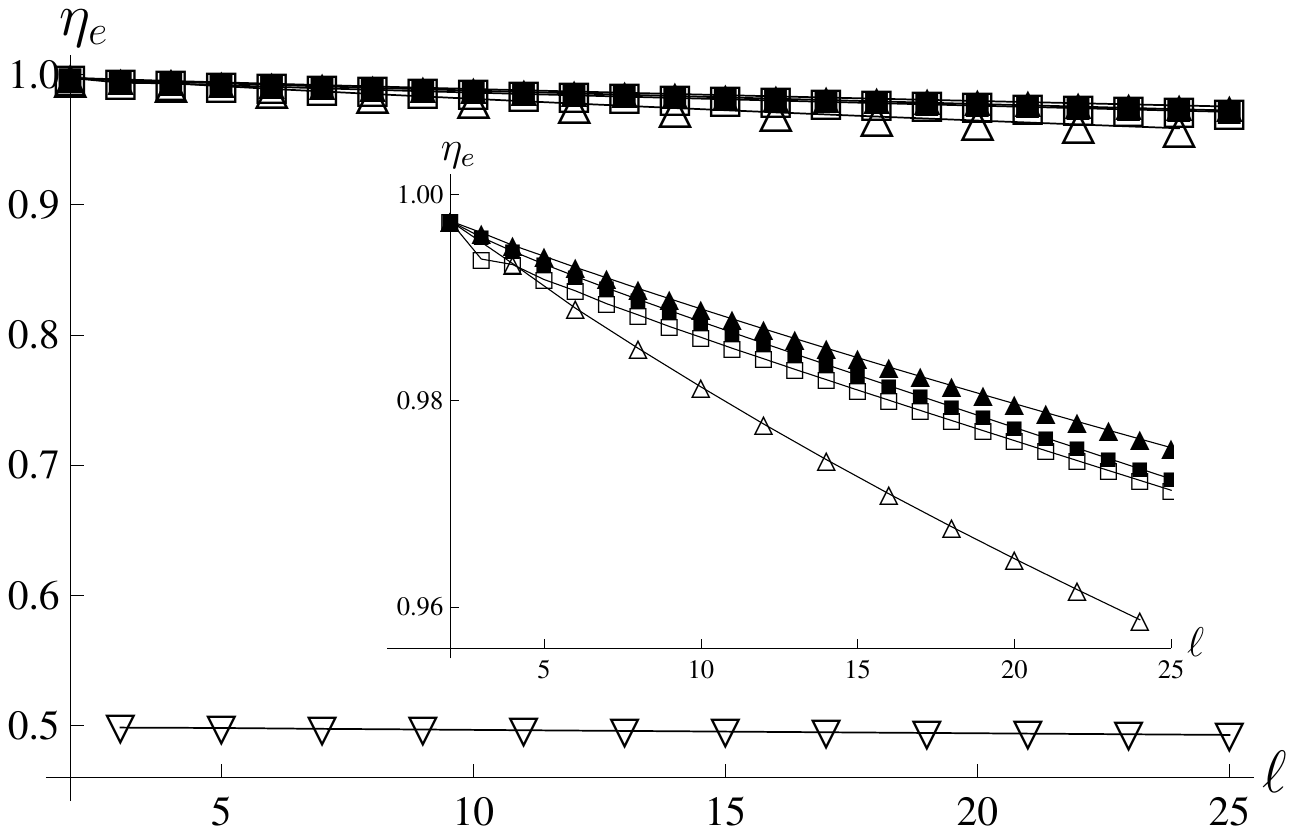}
\caption{Efficiency of energy transfer in the first and second model. Filled (empty) shapes label sink placed at the end (middle) of the chain. Squares (up- and down-triangles) are plots in the first (second) model. Up-triangles ($\vartriangle$) and down-triangles ($\triangledown$) label chains with even and odd sites, respectively.
}
\label{fig4}
\end{figure}

\hfill

\noindent \textit{(1) First model}

\hfill

\Fig{fig4} shows that the energy transfer efficiency using the parameters listed in the first paragraph of the last section. In the first model at low source power $s=0.001\,\psi$, $\blacksquare$- and $\square$-curve label a sink prepared at the end and at the middle of the chain, respectively.
Though the efficiency of the former configuration is slightly greater than the latter, they almost coincide, see the inset of \Fig{fig4}.
At high source power, we already learned from Ref.~\cite{Tay21} that positioning a sink at the middle yields better efficiency (not shown in the figure). This is because at high source power, excitations will accumulate to the lowest exciton level \cite{Frohlich68,Frohlich68b}.
The excitation profile of the lowest exciton level in the site basis turns out to have a maximum at the middle of the chain, as illustrated by the $\bullet$-curve in \Fig{fig2}(a). As a result, a sink placed at the middle is more efficient in trapping energy from the chain.

\hfill

\noindent \textit{(2) Second model}

\hfill

\begin{figure}[t]
\centering
\includegraphics[width=3.6in,trim={4cm 10cm 2cm 11cm}]{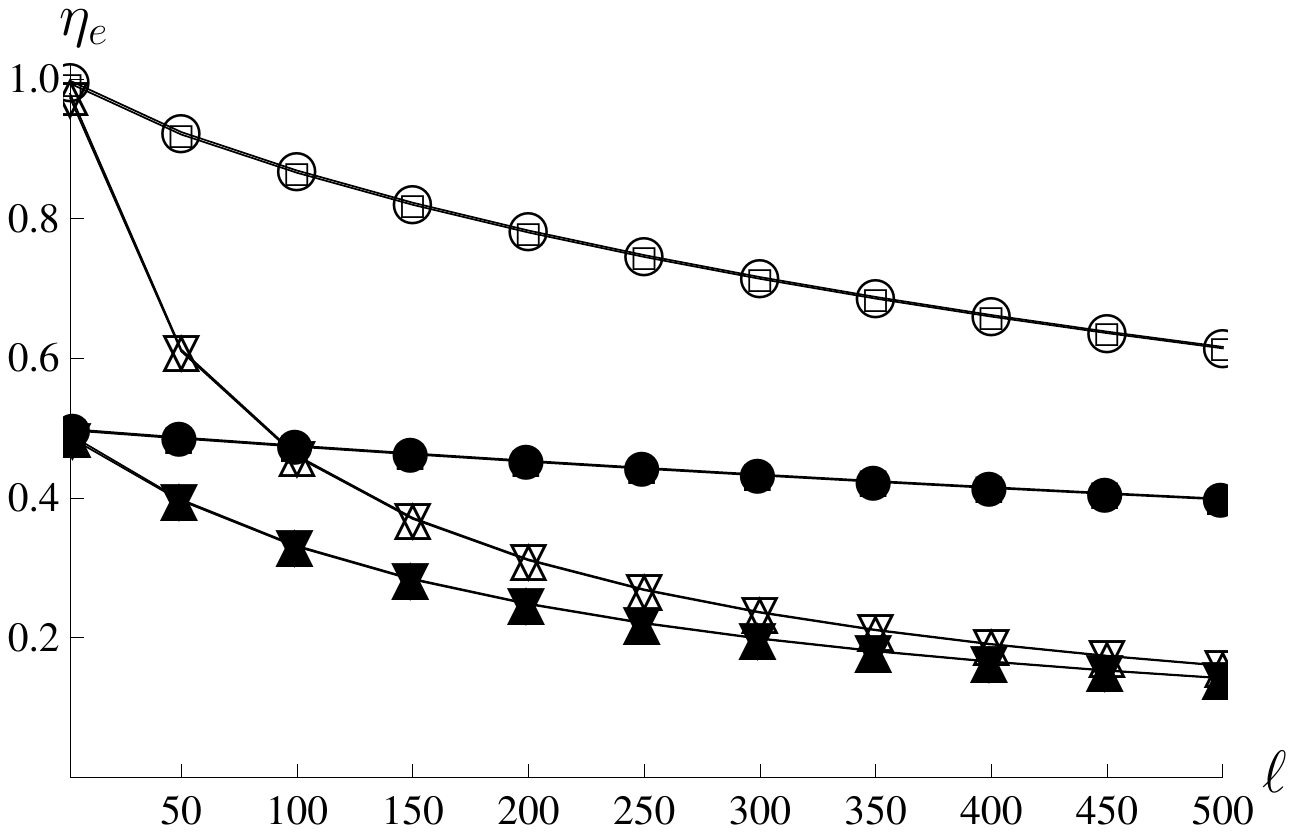}
\caption{
Energy transfer efficiencies in the second model with sinks located at the middle of the chain at different sink rate $\gam_\si$.
The plots show that (1) efficiencies are independent of source power, and (2) efficiencies of the even-$\ell$ chain approach those of the odd-$\ell$ chain in the long chain limit.
Filled and empty shapes label chains with odd number of sites $\ell=3, 49, 99, 149, \cdots$, and even number of sites $\ell=2, 50, 100, 150, \cdots$, respectively.
For $\gam_\s=1\,\psi$, $(\bullet, \circ)$ and $(\blacksquare, \square)$ label a source power of $s=0.001\,\psi$ and $10\,\psi$, respectively. For $\gam_\s=0.1\,\psi$, $\triangledown$ and $\vartriangle$ label a source power of $s=0.001\,\psi$ and $10\,\psi$, respectively. Other parameters are listed in the first paragraph of the previous section.
}
\label{fig5}
\end{figure}

In the second model, a sink prepared at the end of the chain ($\blacktriangle$-curve) gives very similar efficiency as the first model, see the inset in \Fig{fig4}.
However, when the position of the sink shifts to the middle of the chain, the efficiency becomes smaller and depends on whether the chain has odd or even number of sites.
The efficiency is almost halved in odd-$\ell$ chain ($\triangledown$-curve) compared to even-$\ell$ chain ($\vartriangle$-curve).
This is dictated by the coupled-in-pair excitation structure of the second model, as well as by the distribution in the power of the sink across the exciton modes.
In fact, half of the sink power $\al^{(z)}_\sig\gam_\si$ vanishes in odd-$\ell$ chain for a sink at the middle $z=(\ell+1)/2$.
It is interesting to note that this reveals the discrete nature of the chain in that it distinguishes between odd-$\ell$ chain that is less efficient in transferring energy to the sink compared to even-$\ell$ chain.

We expect that the discrete nature becomes less prominent in very long chain.
In fact, when the site number increases, the efficiency in even-$\ell$ chain approaches the one in odd-$\ell$ chain. This is illustrated in \Fig{fig5}, which shows the efficiencies for two values of sink power $\gam_\s$ when energy is supplied at two different source power.
The rest of the parameters used in the plots are listed in the first paragraph of the previous section.
In the figure, filled shapes $(\bullet, \blacksquare, \blacktriangledown, \blacktriangle)$ denote chains with odd $\ell=3, 49, 99, 149, \cdots$, and empty shapes $(\circ, \square, \triangledown, \vartriangle)$ denote chain with even $\ell=2, 50, 100, 150, \cdots$. The sinks are located at the middle of the chain in all the plots. At a rate of sink $\gam_\s=1\,\psi$, $(\bullet, \circ)$ and $(\blacksquare, \square)$ label $s=0.001\,\psi$ and $10\,\psi$, respectively. Whereas when $\gam_\s$ is reduced to $0.1\,\psi$, $(\blacktriangledown,\triangledown)$ and $(\blacktriangle,\vartriangle)$ label $s=0.001\,\psi$ and $10\,\psi$, respectively.

The plots verify the fact that in the second model, the energy transfer efficiency is independent of the source power, a result that follows from \Eq{nsite2}.
We also notice that in all cases the efficiency of even-$\ell$ chain approaches the one of odd-$\ell$ chain as $\ell$ increases. The approach is faster for smaller sink power $\gam_s$. In very long chain, the efficiency should be insensitive to whether the chain has odd or even number of sites.

\section{Conclusion}

We show that by designing different models of coupling between the oscillators and the phonon bath in a quantum molecular chain, it is possible to produce different excitation profiles along the chain with different efficiencies.
When the difference in the site-phonon coupling alternates in sign between successive sites along a chain, the structure in the exciton couplings is very different compared to a model introduced in our previous work, where the difference in the site-phonon coupling is the same in successive sites. In the former model, only pairs of excitons are coupled, and all transitions have the same relaxation rate. Furthermore, the excitation profiles in the site basis are independent of the source power. It is also interesting to learn that a sink prepared at the middle of the chain is able to tell apart whether the chain has odd or even number of sites in that they yield very different efficiency. As the number of sites increases, the evenness or oddness of the chain becomes less obvious and the efficiency approaches each other.

\begin{acknowledgments}
We thank Professor Kurunathan Ratnavelu for inviting us to give a talk in the 1st International Conference on Computational Science \& Data Analytics (COMDATA 2021), 21-24 Nov 2021.
This work is supported by the Ministry of Higher Education, Malaysia (MOHE) under the Fundamental Research Grant Scheme (FRGS), Grant No.~FRGS/1/2020/STG07/UNIM/02/1.
\end{acknowledgments}

\providecommand{\noopsort}[1]{}\providecommand{\singleletter}[1]{#1}%

\end{document}